\documentclass[11pt]{article} 
\newcommand{\be}{\begin{equation}} 
\newcommand{\ee}{\end{equation}} 
\newcommand{\bea}{\begin{eqnarray}} 
\newcommand{\eea}{\end{eqnarray}} 
\newcommand{\bean}{\begin{eqnarray*}} 
\newcommand{\eean}{\end{eqnarray*}} 
\newcommand{\brray}{\begin{array}} 
\newcommand{\erray}{\end{array}} 
\newcommand{\ben}{\begin{equation}{nonumber}} 
\newcommand{\een}{\end{equation}{nonumber}}

\newtheorem{dfn}{Definition}[section] 
\newtheorem{thm}[dfn]{Theorem} 
\newtheorem{lmma}[dfn]{Lemma} 
\newtheorem{ppsn}[dfn]{Proposition} 
\newtheorem{crlre}[dfn]{Corollary} 
\newtheorem{xmpl}[dfn]{Example} 
\newtheorem{rmrk}[dfn]{Remark}

\newcommand{\bdfn}{\begin{dfn}} 
\newcommand{\bthm}{\begin{thm}} 
\newcommand{\blmma}{\begin{lmma}} 
\newcommand{\bppsn}{\begin{ppsn}} 
\newcommand{\bcrlre}{\begin{crlre}} 
\newcommand{\bxmpl}{\begin{xmpl}} 
\newcommand{\brmrk}{\begin{rmrk}} 
 
\newcommand{\edfn}{\end{dfn}} 
\newcommand{\ethm}{\end{thm}} 
\newcommand{\elmma}{\end{lmma}} 
\newcommand{\eppsn}{\end{ppsn}} 
\newcommand{\ecrlre}{\end{crlre}} 
\newcommand{\exmpl}{\end{xmpl}} 
\newcommand{\ermrk}{\end{rmrk}} 
 
\newcommand{\cla}{{\cal A}} 
\newcommand{\clb}{{\cal B}}

\newcommand{\clh}{{\cal H}}

\newcommand{\clk}{{\cal K}} 
\newcommand{\cll}{{\cal L}}

\newcommand{\cls}{{\cal S}}

\def\a*{{\cal A}_{h,*}} 
\def\B{{\cal B}(h)} 
\def\B1{{\cal B}_1(h)} 
\def\b{{\cal B}^{s. a. }(h)} 
\def\b1{{\cal B}^{s. a. }_1(h)}

\newcommand{\ot}{\otimes}

\newcommand{\raro}{\rightarrow}

\newcommand{\lgl}{\langle} 
\newcommand{\rgl}{\rangle}

\begin{document} 

 
\begin{center} 
{\bf{\large Twisted entire cyclic cohomology, J-L-O cocycles and equivariant
spectral triples}}\\  
by\\  Debashish Goswami \\ \emph{The Abdus Salam International Centre For
Theoretical Physics (Mathematics Section),\\ Strada Costiera 11, 34014
Trieste, Italy;  \\   email 
: goswamid@ictp.trieste.it} 
\end{center} 
\begin{abstract}
We study the ``quantized calculus" corresponding to the algebraic ideas
related to ``twisted cyclic cohomology" introduced in \cite{KMT}. With very
similar definitions and techniques as those used in \cite{jlo}, we define and
study ``twisted entire cyclic cohomology" and the ``twisted Chern character"
associated with an appropriate operator theoretic data called ``twisted
spectral data", which consists of a spectral triple in the conventional sense
of noncommutative geometry (\cite{Con}) and an additional positive
operator having some specified properties. Furthermore, it is shown that given
a spectral triple (in the conventional sense) which is equivariant under the
action of a compact matrix pseudogroup, it is possible to obtain a canonical
twisted  spectral data and hence the corresponding (twisted) Chern character,
which will be invariant under the action of the pseudogroup, in contrast to
the fact that the Chern character coming from the conventional noncommutative
geometry need not be invariant under the above action.       
  \end{abstract}

\section{Introduction}
Ordinary and entire cyclic cohomology theory are indeed some of the
fundamental ingredients of Connes' noncommutative geometry. A comprehensive
account of this theory can be found in \cite{Con} and the references cited
 in that book. Let us briefly recall how this theory is used to define a
noncommutative version of the Chern character. First of all, there is a
canonical pairing between the K-theory and the ordinary as well as the entire
cyclic cohomology. Let $\cla$ be a Banach or more generally locally
convex topological algebra, $\lgl .,. \rgl : K_*(\cla) \times H^*(\cla) \raro
C$ and $\lgl .,. \rgl_\epsilon : K_*(\cla) \times H^*_\epsilon (\cla) \raro C$
be the canonical pairing (c.f. \cite{Con}) between the K-theory and the
periodic cyclic cohomology and the pairing between the K-theory and the entire
cyclic cohomology of $\cla$ respectively. Given a Fredholm module $(\clh,
F)$ over $\cla$, or equivalently a spectral triple $(\cla, \clh, D)$, one
constructs a canonical element $ch_*(\clh, F)$, called the chern character, of
$H^*(\cla)$ or $H^*_\epsilon(\cla)$, depending on whether the Fredholm module
is ``$p$-summable" for some $p>0$ or it is only ``$\Theta$-summable", and
$\ast$ will stand for even or odd depending on whether the Fredholm module is
even or odd (i.e. equipped with a compatible grading or not). The map $\phi :
 K_*(\cla) \raro C$ given by $\phi(.)=\lgl, ch_*(\clh,F) \rgl$ ( $\lgl,
ch_*(\clh,F) \rgl_\epsilon$ for the $\Theta$-summable case) actually takes
integer values, and can be obtained as an index of a suitable Fredholm
operator. 

For the finite summable situation, $ch_*(\clh,F)$ is given by (upto a
constant) the cocycle $\tau_n(a_0,...a_n)=Tr_s(a_0[F,a_1]...[F,a_n]), a_j \in
\cla$, where $Tr_s$ is a kind of graded trace defined in \cite{Con}. In terms
of the associated spectral triple, under some additional assumption, one gets
a canonical Hochschild $n$-cocycle $\phi_w$ given by, $\phi_w(a_0,..a_n)=
 \lambda_n \Psi(a_0[D,a_1]...[D,a_n])$, where $\lambda_n$ is a constant and
$\Psi(A):=Tr_w(A |D|^{-n}), A \in \clb(\clh)$, where $Tr_w(.)$ denotes the
Dixmier trace and $n$ is suitably chosen to have $|D|^{-n}$ in the
Dixmier-trace-class. The positive linear functional $\cla \in a \mapsto
Tr_w(a|D|^{-n})$ is a trace and can be thought of as the ``noncommutative
volume form" associated with the noncomutative spin geometry encoded by the
spectral triple. Furthermore, if $\cla$ is equipped with the action of a
classical compact group $G$ coming from a unitary representation on $\clh$ and
the spectral triple is   $G$-equivariant (which means in particular that $D$
commutes with the $G$-representation on $\clh$), then the above-mentioned
 cocycles are $G$-invariant in the sense that
$\tau_n(g.a_0,....g.a_n)=\tau(a_0,..a_n)$ and similar thing is true for
$\phi_w$. In particular, if $\cla$ is chosen to be an appropriate function
algebra (containing the smooth functions) on a classical compact Lie group
$G$ and the canonical equivariant Dirac operator $D$ is chosen, then the
above-mentioned volume form will (upto a constant) coincide with the integral
with respect to the Haar measure. The above-mentioned invariance of the  
cocycles makes it possible to consider its lifting to the algebraic
crossed-product in a canonical way.

However, things do drastically change when one replaces classical compact
groups by noncommutaive and non-cocommutative comapct quantum groups defined by
Woronowicz (\cite{Wo1}). The first major difference is that the canonical Haar
state on such quantum groups are no longer tracial, and if one considers an
equivariant spectral triple such as the ones constructed by Chakraborty and
Pal (c.f. \cite{CP}), and constructs the Chern character as mentioned
before,  it will no longer be invariant under the natural action of the
quantum group.    In
particular, the noncomutative volume form in these cases will not coincide
with the Haar state, and in fact need not be even faithful. The simplest and
important case of $SU_q(2)$ ($0<q<1$) deserves some discussion in this
context. From the explicit description of the $K$-homology of $SU_q(2)$ in
\cite{MNW}, it is easily seen that the Chern character $\omega$ (in the
notation of the above-mentioned paper) of the $1$-summable  generator of the
odd $K$-homology is not invariant under the $SU_q(2)$-action. Had it been
invariant, one would have $\omega \ast \tau_{even} =\omega,$ which is not the
case as it is shown in \cite{MNW} 
 (here $\tau_{even}$ is as in \cite{MNW} and $\ast$ denotes the product
described in that paper, which is the combination of the shuffle product and
the $SU_q(2)$-coproduct). Thus, it is not possible to get an invariant Chern
character within the framework of conventional noncommutative geometry, and
this explians why the chern character obtained in \cite{CP} cannot be
invariant. However, even if one forgets the aspect of invariance, there are
other strange properties observed in this example. It can be shown (to be
discussed in detail elsewhere) that some of the natural properties which one
almost always observes for a nice compact manifold such as the classcical
$SU(2)$ are not valid. For example, there can be no $1$-summable odd Dirac
operator $D_1$ for $SU_q(2)$ whose associatd Chern character is nontrivial and
 which also satisfies the property that repeated commutators with $|D_1|$ is
bounded for any element in the finite $\ast$-algebraic span of the canonical
generators of $SU_q(2).$  It is interesting to remark that the equivariant
Dirac operator in \cite{CP}, whose associated Fredholm module is the generator
of the odd $K$-homology group, is $3$-summable and the Hochschild cohomology
class of the associated Chern character vanishes, thus this Chern character
(which is in $HC^3$) must be of the form $S\tau$ for some $\tau \in HC^1 $
(where $S$ is the periodicity operator used in \cite{Con} in the exact couple
relation between the Hochschild and cyclic cohomology), and it is not clear
whether it is at all possible to obtain this $\tau$ as a Chern character
coming from some nice equivariant $1$-summable Dirac operator. Furthermore, as
will be discussed elsewhere (\cite{Gos2}), it is far from clear  (due to some
more strange properties ) whether  it is possible to describe the Chern
character by a ``local formula", in the sense of \cite{Conloc}. In any case, 
all these facts suggest that there may be some incompatibility between the 
existing framework of noncommutative geometry and the theory of compact
quantum groups, since even for the simplest such quantum group a few anomalies
seem to occur, specially in context of the ``local" formulae. We feel that the
research in this direction (i.e. towards finding a framework in which both
noncommutative manifolds and quantum groups are well fitted)  is still at a
beginning or rather somewhat experimental stage, and it will take more time to
reach a conclusive answer. Thus, at this moment, it may be worthwhile to
explore some alternative frameworks of noncommutative geometry too, and try to
compare the relative advantages and disadvantages of different approaches in
the light of various examples at hand.  In the present article, we shall focus
our attention on one such alternative approach suggested in \cite{KMT}, based
on what the authors have called ``twisted cyclic cohomology". We shall study
the operator theoretic framework for that, and for some natural reason deal
with its ``entire" version. We would like to point out here that we shall build
some amount of general theory only, which in particular will enable one to
obtain an invariant (twisted) Chern character in this context, but we shall
leave  the study of particular examples for future. Thus, apart from the
aspect of invariance, we do not know yet whether this alternative framework  
will help us to understand $SU_q(2)$ and similar models better than the
conventional theory; but we hope to take up that issue later on.         

Motivated by the fact that the Haar state on typical compact quantum groups
are not tracial and other things,  the authors of \cite{KMT} have found it
somewhat natural to introduce ``twisted cyclic cohomology", which is indeed a
module in the cyclic category (see. e.g. \cite{Con} or \cite{Loday}). However,
they did not focuss on the ``quantized calculus" related the twisted cyclic
cohomology, which is our goal in the present article. To be more precise, we
shall discuss the twisted analogue of the entire cyclic cohomology and show
how one can obtain canonical J-L-O type {c.f. \cite{jlo}) cocycles in this
twisted entire cyclic cohomology from a spectral triple and an additional
positive operator giving rise to the ``twist". In fact, although we shall make
a somewhat general theory, our main focuss will be on the examples coming from
the quantum group theory and we shall show that a canonical ``twisting"
operator exists for a given equivariant spectral triple for the action of
compact matrix pseudogroups of Woronowicz. Let us remark here that in some
special examples of noncommutative manifolds studied in \cite{CD} and
\cite{CL}, the conventional theory of noncommutative geometry was shown to be
nicely applicable to certain Hopf algebras or associated homogeneous spaces,
but those Hopf algebras (e.g. $SL_q(2)$ for $q$ complex of modulus $1$) do not
come under the framework  of topological quantum groups given by Woronowicz
and others. 

Before we enter into the main results, we should perhaps mention why we are
interested in the twisted version of the entire cyclic cohomolgy (hence J-L-O
type cocycles) rather than the ordinary cyclic cohomology. This is motivated
by our study of the example $SU_q(2)$ (\cite{Gos}, and also \cite{CP}), where
we have shown that  the Haar state can be recaptured by the formula
$\frac{Tr(aR e^{-tD^2}) }{Tr(Re^{-tD^2})}, a \in SU_q(2)$, where $D$ is the
equivariant Dirac operator and $R$ is a suitable positive operator, coming
from the modular theory of $SU_q(2)$. It is also shown that there is no finite
postiive number $d$ so that $Tr(Re^{-tD^2})=O(t^{-d})$. This in some sense
indicates that the associated Fredholm module is not finite dimensioanl, or in
other words in $\Theta$-summable, so that it is natural to construct $J-L-O$
type cocyles in the (twisted) entire cyclic cohomology.  
\section{Twisted entire cyclic cohomology and J-L-O cycles}
\subsection{Algebraic aspects}
Let us  develop the theory for Banach algebras for simplicity, but we note
that  our results  extend  to locally convex algebras, which we
actually need. The extnesion to the locally convex algebra case follows
exactly as remarked in [1, page 370]. So, let $\cla$ be a unital
Banach algebra, with $\| . \|_*$ denoting its Banach norm, and let $\sigma$ be
a continuous automorphism of $\cla$, $\sigma(1)=1$. For $n \geq 0$, let $C^n$
be the space of continuous $n+1$-linear functionals $\phi$ on $\cla$ which are
$\sigma$-invariant, i.e. $\phi(\sigma(a_0),..\sigma(a_n))=\phi(a_0,...a_n)
\forall a_0,...a_n \in \cla$; and $C^n =\{0\}$ for $n<0$. We define linear
maps $T_n, N_n : C^n \raro C^n$,  $ U_n : C^n \raro C^{n-1}$ and $V_n : C^n
\raro C^{n+1}$ by, $$ (T_nf)(a_0,...a_n)=(-1)^n f(\sigma(a_n),
a_0,...a_{n-1}), N_n=\sum_{j=0}^n T_n^j,$$
$$ (U_nf)(a_0,...a_{n-1})=(-1)^n f(a_0,...,a_{n-1}, 1),$$
$$(V_nf)(a_0,...a_{n+1})=(-1)^{n+1} f(\sigma(a_{n+1})a_0,a_1,...,a_n).$$
 Let $B_n=N_{n-1}U_n (T_n-I),$ $b_n=\sum_{j=0}^{n+1} T_{n+1}^{-j-1}V_nT_n^j$. 
  Let $B,b$ be maps on the complex $C\equiv (C^n)n$ given by $B|_{C^n}=B_n,
b|_{C^n}=b_n$. It is easy to verify (similar to what is done for the untwisted
case , e.g. in \cite{Con}) that $B^2=0$, $b^2=0$ and $Bb=-bB$, so that we get
a bicomplex $(C^{n,m}\equiv C^{n-m})$ with differentials $d_1,d_2$ given by
$d_1=(n-m+1)b : C^{n,m} \raro C^{n+1,m},$ $d_2=\frac{B}{n-m} : C^{n,m} \raro
C^{n,m+1}.$ Furthermore, let $C^{e}=\{ (\phi_{2n}){n \in N}; \phi_{2n} \in
C^{2n} \forall n \in N \},$ and $C^o=\{ ( \phi_{2n+1}){n \in N}; \phi_{2n+1}
\in C^{2n+1} \forall n \in N \}$. We say that an element $\phi=(\phi_{2n})$ of
$C^e$ is a $\sigma$-twisted even entire cochain if the radius of convergence of
the complex power series $\sum \| \phi_{2n} \|\frac{z^n}{n!}$ is infinity,
where $\| \phi_{2n} \| := \sup_{ \| a_j \|_* \leq 1} | \phi_{2n}
(a_0,....,a_{2n})|.$ Similarly we define $\sigma$-twisted odd entire cochains,
and let $C^e_\epsilon(\cla, \sigma)  $ ($C^o(\cla,\sigma)$ respectively) denote
the set of $\sigma$-twisted even (respectively odd) entire cochains. Let
$\partial =d_1+d_2$ , and we have the shrot complex $C^e_\epsilon(\cla,
\sigma)
\stackrel{\stackrel{\partial}
{\longleftarrow}}{\stackrel{\longrightarrow}{\partial}}  C^o_\epsilon(\cla,
\sigma)$. We call the cohomology of this complex the $\sigma$-twisted entire
cyclic cohomology of $\cla$ and denote it by $H^*_\epsilon(\cla,\sigma)$. 

\bppsn
\label{pair}
Let $\cla_\sigma=\{ a \in \cla : \sigma(a)=a \}$ be the fixed point subalgebra
for the automorphism $\sigma$. There is a canonical pairing
$<.,.>_{\sigma,\epsilon} : K_*(\cla_\sigma) \times H^*_\epsilon(\cla, \sigma)
\raro C$.
\eppsn
The proof is omitted, since it is very similar to the untwisted case, for
example, as given in \cite{Con}. In fact, this pairing is nothing but the
pairing between the $K$-theory of $\cla_\sigma$ and the entire cyclic
(untwisted) cohomologies of $\cla_\sigma$, as any element in the
$\sigma$-twisted entire cyclic cohomology of $\cla$ can be viewed as an
(untwisted) entire cyclic cocycle on $\cla_\sigma$ by restriction on
$\cla_\sigma.$ Thus, the arguments for the untwisted case apply to our
situation to prove the above proposition. 

\subsection{Construction of the Chern character using the J-L-O cocycles}
We begin with the following definition :\\
\bdfn
\label{spectral}
Let $\clh$ be a separable Hilbert space, $\cla^\infty$ be a $subalgebra$ (not
necessarily complete) of $\clb(\clh)$, $R$ be a positive (possibly unbounded)
operator in $\clh$, $D$ be a self-adjoint operator in $\clh$ such that the
followings hold :\\
(i) $[D,a] \in \clb(\clh) \forall a \in \cla^\infty$,\\
(ii) $R$ commutes with $D$,\\
(iii) For any real number $s$ and $a \in \cla^\infty$,
$\sigma_s(a):=R^{-s}aR^s$  is bounded and belongs to $\cla^\infty$.
Furthermore, $a \mapsto \sigma_s(a) $ is an automorphism of $\cla^\infty$ and 
 for any positive integer $n$, $\sup_{s \in [-n,n]} \| \sigma_s(a) \| <
\infty.$\\
Then we call the quadruplet $(\cla^\infty, \clh, D, R)$ an odd $R$-twisted
spectral data. Furthermore, if there is a grading given by $\gamma \in
\clb(\clh)$ with $\gamma=\gamma^*=\gamma^{-1}$, and $\gamma$ commutes with
$\cla^\infty$ and $R$, and anticommutes with $D$,
 then we say that we are given an even $R$-twisted spectral data. We say that
the given (odd or even) twisted spectral data is $\Theta$-summable if 
$Re^{-tD^2}$ is trace-class for all $t >0$. 
\edfn    

Let us consider the case of even twisted spectral data only, as the odd case
can be treated with obvious and minor modifications as done in the untwisted
case. Let us assume that we are given such an even twisted spectral data
specified by $\cla^\infty, \clh, D, R, \gamma$ as in the above definition, and
 fix any $\beta >0$. 
 Let $H=D^2$, $A^{(s)}=e^{-sH}Ae^{-sH},$ $A(s)=e^{-sH}Ae^{sH}$ for $s >0$ and
$A \in \clb(\clh).$ Let us denote by $\clb$ the set of all $A \in \clb(\clh)$
for which $\sigma_s(A):=R^{-s}AR^s \in \clb(\clh) $ for all real number $s$,
$[D,A] \in \clb(\clh)$ and $s \mapsto \| \sigma_s(A) \|$ is bounded over
compact subsets of the real line.  We define for $n \in N$ an $n+1$-linear
functional $F_n^\beta$ on $\clb$ by the formula $$
F^\beta_n(A_0,...A_n)=\int_{\sigma_n}Tr(\gamma A_0
A_1(t_1)..A_n(t_n)Re^{-\beta H})dt_1...dt_n,$$  where $\sigma_n =\{
(t_1,...t_n) : 0 \leq t_1 \leq ...\leq t_n \leq \beta \}$. That the above
integral exists as a finite quantity follows from the following lemma. \blmma 
$F_n^\beta$ is well defined and one has the estimate  
$$ |F^\beta_n(A_0,..A_n)| \leq \frac{\beta^n}{n!} Tr(Re^{-\beta H})
\Pi_{j=o}^n C_j,$$ where $C_j=\sup_{s \in [-1,1] } \| \sigma_s(A_j) \|$. 
\elmma
{\it Proof :-}\\
The proof is very similar to that of Proposition IV.2 pf \cite{jlo}, so we
only sketch the main ideas. We use the same notation
$\delta_1,...\delta_{n+1}$ as in \cite{jlo}, i.e.
$\delta_j=\frac{t_j-t_{j-1}}{\beta}$, with $t_0=0, t_{n+1}=\beta$. Thus,
$(t_1,..t_n)$ in the integrand can be replaced by
$(\delta_1,......\delta_{n+1})$ with the condition that $\delta_j \geq 0, \sum
\delta_j =1$. Then, as in \cite{jlo}, we have that 
\bean
\lefteqn {Tr (
A_0A_1(t_1)..A_n(t_n)Re^{-\beta H})}\\
& =&
Tr \left( \gamma A_0 e^{-\beta \delta_1H}A_1
e^{-\beta \delta_2 H}...A_n e^{-\beta \delta_{n+1}H} R^{\sum_j
\delta_j} \right)\\
&=& Tr \left( \gamma \sigma^{-1}(A_0)(Re^{-\beta H})^{\delta_1}
\sigma^{-\sum_{j=2}^{n+1} \delta_j}(A_1)(Re^{-\beta
H})^{\delta_2}....\sigma^{-\delta_{n+1}}(A_n)(Re^{-\beta H})^{\delta_{n+1}}
\right),\\
\eean
where in the last step we have used the fact that $R$ and $H$ commute, and
$\gamma$ and $R$ also commute. Now, by the 
  generalised Holder's inequality for Schetten ideals the desired estimate
follows, noting that $\|(Re^{-\beta H})^{\delta}
\|_{\delta^{-1}}=Tr(Re^{-\beta H})^{\delta}$, as $Re^{-\beta H}$ is a positive 
operator.\\

 For $A \in \clb(\clh)$, let $\dot{A} $ denote
$\frac{d}{dt}(A(t))|_{t=0}=-[H,A]$, whenever it exists as a bounded operator.
Clearly for $A$ of the form $A=B^{(s)}, s>0$, $\dot{A} \in \clb(\clh)$. 
\blmma
\label{formula}
Let $A_i,i=0,1,...,n$ be elements of $\clb$ such that $\dot{A_i } \in
\clb(\clh) \forall i$. Let $dA:=i[D,A]$. Then we have the following :\\
(i) For $j=1,...,n,$ 
$F^\beta_{n+1}(A_0,...,A_{j-1},\dot{A_j},...,A_{n+1})=F^\beta_n(A_0,...,A_{j-1},A_jA_{j+1},...,A_{n+1})-F^\beta_n(A_0,...A_{j-1}A_j,...,A_{n+1}).$\\ 
(ii) 
$F^\beta_{n+1}(\dot{A_0},A_1,...,A_{n+1})=F^\beta_n(A_0A_1,...,
A_{n+1})-F^\beta_n(A_1,...,A_{n+1}\sigma_{-1}(A_0)).$\\ 
(iii)
$F^\beta_{n+1}(A_0,A_1,...,\dot{A_{n+1}})=F^\beta_n(\sigma_1(A_{n+1})A_0,...,
A_{n})-F^\beta_n(A_0,A_1,...,A_nA_{n+1}).$\\ 
(iv) $F^\beta_n(A_0,...,A_n)=F^\beta_n(\sigma_1(A_n),A_0,...,A_{n-1})$ and
$F^\beta_n(\sigma_1(A_0),...,\sigma_1(A_n))=F^\beta_n(A_0,...,A_n).$\\
(v) $\sum_{j=0}^{n-1} F^\beta_n(A_0,...,A_j,1,A_{j+1},...,A_{n-1})=\beta
F^\beta_{n-1}(A_0,...,A_{n-1}).$\\
(vi) $F^\beta_n(dA_0,...,dA_n)=\sum_{j=1}^n (-1)^jF^\beta_n
(A_0,dA_1,...,dA_{j-1},\dot{A_j},dA_{j+1},...,dA_n).$
\elmma
The proof of the above formulae are straightforward and very similar to the
analogous formulae derived in \cite{jlo}, hence we omit the proof. 

Let us now equip $\cla^\infty$ with the locally convex topology given by the
family of Banach norms $\|.\|_{*,n},n=1,2,...$, where $\|a\|_{*,n}:= \sup_{s
\in [-n,n]} ( \| \sigma_s(a) \|+\| [D,\sigma_s(a)] \|)$. Let $\cla$ denote the
completion of $\cla^\infty$ under this topology, and thus $\cla$ is Frechet
space. We shall now construct the Chern character in
$H^*_\epsilon(\cla,\sigma)$, where $\sigma=\sigma_1$, which extends on the
whole of $\cla$ by continuity. 
\bthm
\label{chern}
Let $\phi^e\equiv (\phi_{2n})$ and $\phi^o\equiv (\phi_{2n+1})$ be defined by
$$
\phi_{2n}(a_0,...,a_{2n})=\beta^{-n}F^\beta_{2n}(a_0,[D,a_1],...,[D,a_{2n}]),
a_i \in \cla,$$ 
$$ \phi_{2n+1}(a_0,...,a_{2n+1})=\sqrt{2i}
\beta^{-n-\frac{1}{2}}F_{2n+1}^\beta(\gamma a_0,[D,a_1],...,[D,a_{2n+1}]), a_i
\in \cla.$$ Then $(b+B)\phi^e=0$, $(b+B)\phi^o=0$, and hence $\psi^e\equiv
((2n)!\phi_{2n}) \in H^e_\epsilon(\cla,\sigma)$ and $\psi^0 \equiv ((2n+1)!
\phi_{2n+1}) \in H^o_\epsilon(\cla,\sigma).$
\ethm
{\it Proof :-}\\
First we extend the definition of $\phi_{2n}, \phi_{2n+1}$ on the whole of
multiple copies of  $\clb$ by the same formula, which is clearly well defined.
 Let $\clb^\infty$ denote the unital 
algebraic span of elements of the  form $A^{(s)}$ for $s>0$ and $A \in
\cla^\infty$. Let us denote by $C^n(\clb^\infty)$ the space of all
$n+1$-linear functionals on $\clb^\infty$ (without any continuity
requirements) and extend the definitions of $b$ and $B$ on the complex
$C(\clb^\infty)\equiv (C^n(\clb^\infty)_n)$ by the same expression as in the
case of $C^n$, i.e. for functionals on $\cla$. This is possible because
$\sigma=\sigma_1$ is defined on the whole of $\clb^\infty$. Now, the formulae
(i) to (vi) of Lemma (\ref{formula}) are applicable for elements of
$\clb^\infty$, and by a straightforward calculation as in \cite{jlo} we can
show that $(b+B)(\phi^*)=0$ on elements of $\clb^\infty$. Now, to prove the
same for elemenets of $\cla^\infty$, we note that for $A \in \cla^\infty$,
$A^{(s)}
 \raro A,$ $[D,A^{(s)}]=[D,A]^{(s)} \raro [D,A]$ and
$\sigma_t(A^{(s)})\raro \sigma_t(A) \forall t$, as $s \raro 0+$ and the
convergence of operators is w.r.t. the strong operator topology of
$\clb(\clh)$. By using the fact that $Tr(B_nC) \raro Tr(BC)$ if $B_n \raro B$
 w.r.t. the strong operator topology and $C$ is trace-class, we conclude that
the integrand in the definition of $F^{\beta}_{2n}(a_0^{(s)},
[D,a_1^{(s)}],...[D,a_{2n}^{(s)}])$ converges to that with $a_j^{(s)}$
replaced by $a_j$ (for $a_j \in \cla^\infty$), and finally, as $\|a^{(s)}\|
\leq \| a \| \forall a \in \clb(\clh)$, an application of the Dominated
Convergence Theorem allows us prove that $(b+B) \phi^e=0$ on elements of
$\cla^\infty$, and hence by continuity the same thing holds for $\cla$.
Similarly the odd case can be done. The remaining part of the statement of
the theorem is straighforward, and follows exactly in the same way as in
\cite{Con}.

We shall call  $\psi^*$ in the above theorem the {\bf Chern character } of the
twisted spectral data $(\cla^\infty, \clh, D, R)$ (or $(\cla^\infty, \clh, D,
R, \gamma)$ for the even case). We remark here by an easy adaptation of the
  techniques of \cite{jlo} and \cite{jlo2}, we can show that the above chern
characters do not depend on our choice of $\beta$, namely cohomologous for
all $\beta >0$. Furthermore, invariance of the chern character under some
suitable homotopy of the spectral data can possibly be established along the
lines of the above mentioned references. We, however, would like to consider
those issues elsewhere.   

\section{Canonical twisted equivariant spectral data arising from actions of
compact matrix pseudogroups}
In this section we shall show how one can find canonical examples of twisted
spectral data from the theory of compact matrix pseudogroups of Woronowicz
(c.f. \cite{Wo1}. Let $\cls$ be such a compact matrix pseudogroup, with the
matrix elements $t^n_{ij}, n=1,2,..., i,j=1,2,...,d_n$, such that  for each
$n$,   $T_n \equiv (( t^n_{ij}))_{ij=1}^{d_n}$ is a unitary element of
$M_{d_n}(C)\ot \cls$, and the coproduct $\Delta$ and the antipode $\kappa$ are
given by $\Delta(t^n_{ij})=\sum_k t^n_{ik} \ot t^n_{kj}$,
$\kappa(t^n_{ij})=(t^n_{ji})^*$. Let $\clk=L^2(\cls,h)$ be the $GNS$-space
associated to the faithful Haar state $h$ on $\cls$, and we imbed $\cls$ in
$\clb(\clk)$ in the natural manner. We recall from the theory of quantum groups
that a unitary representation of the quantum group $\cls$ is given by a
separable Hilbert space $\clh$ and a unitary element $V$ of $\cll(\clh \ot
\cls) \subseteq \clb(\clh) \ot \clb(\clk)$ with additional properties (c.f.
\cite{Wo2} and other relevant literature on compact quantum groups, and note
that $\cll(\clh \ot \cls)$ denotes the $C^*$-algebra of adjointable linear
maps on the Hilbert module $\clh \ot \cls$), and we can equivalently think of
the representation to be given by   a map $V^\prime$ from $\clh$ to the
Hilbert module $\clh \ot \cls$ given by  $V^\prime(\xi)=V(\xi \ot 1)$, where
$1$ is the identity in $\cls$. For $A \in \clb(\clh)$, we define
$\delta(a)=V(a \ot I)V^* \in \clb(\clh) \ot \clb(\clk)$. Let us assume that
there is a subalgebra $\cla^\infty$ of $\clb(\clh)$ such that
$\delta(\cla^\infty) \subseteq \cla^\infty \ot_{\rm alg} \cls^\infty$, where
$\cls^\infty$ denotes the algebraic span of the  matrix elements of $\cls$ and
their adjoints. Clearly, $\delta : \cla^\infty \raro \cla^\infty \ot_{\rm alg}
\cls^\infty$ is a coaction of the Hopf algebra $\cls^\infty$.  

\blmma
Let $\phi : \cls^\infty \raro C$ is a multiplicative linear functional. We
define a linear map $F_\phi$ on $\clh$ with the domain consisting of all $\xi
\in \clh$ such that $V^\prime \xi \in \clh \ot_{\rm alg} \cls^\infty$,  and
$F_\phi (\xi)=(id \ot \phi)(V^\prime \xi)$ for any $\xi$ in the above domain.
Then we have that\\ (i) $F_\phi$ is densely defined,\\
(ii) For $a \in \cla^\infty$, $a Dom(F_\phi) \subseteq Dom(F_\phi)$ and
$F_\phi(a\xi)=(\phi \ast a)F_\phi(a)$ for $\xi \in Dom(F_\phi),$ where $(\phi
\ast a):=(id \ot \phi)(\delta(a)).$
\elmma
{\it Proof :-}\\
 By the general theory $\clh$ will be decomposed as $\clh = \bigoplus_{n\geq
1,k=1,...,m_n, m_n \leq \infty} \clh_{n,k}$, and there exists an orthonormal
basis $\{ e^{n,k}_j \}_{j=1,...,d_n}$ for $\clh_{n,k}$ such that
$V^\prime e^{n,k}_j=\sum_i e^{n,k}_i \ot t^n_{ij}$. It is obvious that $\clh_n
\equiv \bigoplus_{k} \clh_{n,k} \subseteq Dom(F_\phi)$. This proves (i). For
(ii), we first note that for $\xi \in Dom(F_\phi)$ and $a \in \cla^\infty$, we
have $V^\prime(a\xi)=\delta(a) V^\prime \xi$ (where an element of $\cla^\infty
\ot_{\rm alg} \cls$ is naturally acting from left by multiplication on $\clh
\ot \cls$), which shows that $a \xi \in Dom(F_\phi)$. The fact that $F_\phi(a
\xi)=(\phi \ast a)F_\phi(\xi)$ is verified easily using the multiplicativity
of $\phi$.\\ 

Let us now recall a few facts from the general theory of compact matrix
pseudogroups as in \cite{Wo1}. It is known that for each $n$, there is a unique
$d_n \times d_n$ complex matrix $F_n$ with the following properties :\\
(i) $F_n$ is positive and invertible with $Tr(F_n)=Tr(F_n^{-1})=M_n>0$, say.\\
(ii) If $h$ denotes the  Haar state on $\cls$, then we
have, $h(t^n_{ij} {t^n_{kl}}^*)=\frac{1}{M_n} \delta_{ik} F_n(j,l)$, where
$\delta_{ik}$ is the Kronecker delta.\\ 
(iii) For any complex number $z$, let $\phi_z$ be the functional on
$\cls^\infty$ defined by $\phi_z(t^n_{ij})=(F_n)^z(j,i)$. Then each $\phi_z$
is multiplicative, $(\phi_z \ast 1)=1$ and for any fixed element $a \in
\cls^\infty$, $z \mapsto \phi_z \ast a$ is a complex analytic map.\\

Let us now take $R=F_{\phi_1}$ on $\clh$. With this choice, we have the
following result. Note that we call an $m$-linear functional $\chi$ on
$\cla^\infty$ invariant (or simply invariant if no cinfusion arises) if  we
have $$
\chi({a_1}_{(1)},...,{a_m}_{(1)}){a_1}_{(2)}...{a_m}_{(2)}=\chi(a_1,...,a_m)1_\cls,$$ 
where we have used the Swedler notation  $\delta(a)=a_{(1)} \ot a_{(2)}$,
with summation implied.  \\ 

\bthm
\label{main}
Assume that $(\cla^\infty, \clh,D)$ is an odd equivariant spectral triple in
the sense of \cite{CP}, i.e. $D$ is a  self adjoint operator on $\clh$ 
which is $\cls$-invariant, in the sense that  $V^\prime(Dom(D)) \subseteq
Dom(D) \ot_{\rm alg} \cls^\infty$ and $V^\prime D=(D \ot I)V^\prime$ on
$Dom(D)$,  and 
furthermore, $[D,a] \in \clb(\clh)$ for $a \in \cla^\infty$. Then
$(\cla^\infty, \clh,  D, R)$ is a twisted odd spectral data.
Similarly, if we also have an equivariant grading then we obtain a
twisted even spectral data. Moreover,  if $Re^{-\beta D^2}$ is trace -class for
all  $\beta >0$, then the associated Chern characters are invariant. 
   \ethm
{\it Proof :-}\\
Since the resolvent operators $(i\lambda-D)^{-1}$, $\lambda \in R,$ of $D$ are
equivariant bounded operators, it is clear that $(i \lambda-D)^{-1}
=\bigoplus_n (I \ot B_{n,\lambda})$, where we have identified $\clh_n$, which
is a direct sum of $d_n$-dimensional Hilbert spaces, with $\clh_{n,1} \ot
C^{m_n}$ ($C^\infty :=l^2$), and in this identification we have $B_{n,\lambda}
\in \clb(C^{m_n})$. Thus, $D$ is also of the same form, with $B_{n,\lambda}$
replaced by say $D_n$, which is a self adjoint (possibly unbounded) operator
on $C^{m_n}$. Now, by the definiton of $R$, it is of the from $\bigoplus_n
(r_n |_{\clh_{n,1}} \ot I)$, from which it follows that $R$ and $D$ commute.
Other conditions in the definition of twisted spectral data are verified
easily using the facts about the canonical family of functionals $\phi_z$
noted before. 

 Finally, we shall verify the invariance of the associated chern character.
To this end, first note that if $Re^{-\beta D^2}$ is trace class, then
$e^{-\beta D_n^2}$ is also trace-class for each $n$, hence in particular has a
complete set of eigenvectors in $C^{m_n}$. Thus, by renaming the orthonormal
basis $e^{n,k}_j$ if necessary, we can without loss of generality assume that
$e^{-\beta D_n^2} e^{n,k}_j=\lambda_{n,k} e^{n,k}_j$ and $R e^{n,k}_j=\sum_l
F_n(j,l) e^{n,k}_l$. Let us now use the embedding of  $\cls$  in the
GNS Hilbert space $\clk =L^2(\cls,h)$ associated with the Haar state $h$, and
denote by $1$ the identity of $\cls$ viewed as the canonical cyclic vector in
$\clk$. We have, $h(b)=\lgl 1, b 1 \rgl \forall b \in \cls$. Let $\chi(A)
:=Tr(ARe^{-\beta D^2})$ be the normal positive linear functional on
$\clb(\clh)$. Since the extension of $h$ on  $\clb(\clk)$ given by $h(B)
:=\lgl 1, B 1 \rgl, B \in \clb(\clk)$ is also a positive linear normal
functional, we can define the positive linear normal functional $(\chi \ot h)$
on $\clb(\clh) \ot \clb(\clk)$. We claim that $(\chi \ot h)(V(a \ot 1)
V^*)=\chi(a)$ for any $a \in \clb(\clh)$. The proof of this fact is very
similar to a similar result obtained in the special case of $SU_q(2)$ in
\cite{Gos}. 

{\it Proof of the claim :}\\
We first note that $V^*(e^{n,k}_j \ot 1)=\sum_i e^{n,k}_i \ot (t^n_{ji})^*$.
Thus, we have, 
\bean
\lefteqn{(\chi \ot h)(V(a \ot I)V^*)}\\
&=& \sum_{n,i,j} \lgl e^{n,i}_j \ot 1, V(a \ot I)V^*(Re^{-\beta
D^2}(e^{n,i}_j) \ot 1) \rgl \\
&=& \sum_{n,i,j,r} \lgl V^*(e^{n,i}_j \ot 1), (a \ot
I)V^*(\lambda_{n,i}F_n(j,r)e^{n,i}_r \ot 1) \rgl \\
&=& \sum_{n,i,j,r,k,l} \lambda_{n,i}F_n(j,r) \lgl e^{n,i}_k \ot
(t^n_{jk})^*, (a \ot I)(e^{n,i}_l \ot (t^n_{rl})^*) \rgl \\
&=& \sum_{n,i,j,r,k,l} \lambda_{n,i} F_n(j,r) \lgl e^{n,i}_k, ae^{n,i}_l
\rgl h(t^n_{jk}{t^n_{rl}}^*)\\
&=& \sum_{n,i,j,k,l} \lambda_{n,i} F_n(j,j)  \lgl e^{n,i}_k, ae^{n,i}_l
\rgl \frac{F_n(k,l)}{M_n} \\
&=& \sum_{n,i,k} (\sum_j \frac{F_n(j,j)}{M_n}) \lambda_{n,i} \lgl e^{n,i}_k,
a(\sum_l F_n(k,l) e^{n,i}_l) \rgl \\
&=& \chi(a),\\
\eean
since $M_n=Tr(F_n)$. 
 
Now, it is easy to see that since $V(D \ot I)=(D \ot I)V$ (viewing $V$ as a
unitary in $\clb(\clh) \ot \clb(\clk)$), for $a \in \cla^\infty$ and $s>0$,
one has $V(ae^{-sD^2}\ot I)V^*= \delta(a) (e^{-sD^2} \ot I)$. For $s_0,...,s_n
>0$, let $\eta $ be the $n+1$ linear functional on $\cla^\infty$ given by
 $\eta(a_0,...,a_n)=Tr(a_0e^{-s_0D^2}
[D,a_1]e^{-s_1D^2}...[D,a_n]e^{-s_nD^2}R)$. Clearly, as $V$ commutes with 
 $(D \ot I)$, we have, 
\bean
\lefteqn{V((a_0e^{-s_0D^2}[D,a_1]e^{-s_1 D^2}...[D,a_n]) \ot
I)V^*}\\
&=& \delta(a_0).(e^{-s_0D^2} \ot I).[(D \ot I),\delta(a_1)]...[(D \ot
I),\delta(a_n)]\\
&=&({a_0}_{(1)}e^{-ts_0D^2}...[D,{a_n}_{(1)}]) \ot 
({a_0}_{(2)}...{a_n}{(2)}),\\
\eean 
using the Swedler notation with summation
implied. Hence, by what we have proved earlier, it is easy to see that
$\eta({a_0}_{(1)},...,{a_n}_{(1)})h({a_0}_{(2)}...{a_n}_{(2)})
=\eta(a_0,...,a_n)$. From this the required invariance of the odd chern
characters follows, because by using the fact that $h \ast \lambda =\lambda(1)
h$ for any bounded linear functional $\lambda$ on $\cls$ (where $\ast$ is the
convolution product of two linear functionals on $\cls$, defined for example
in \cite{Wo1}), we get that $
\eta({a_0}_{(1)},...,{a_n}_{(1)})\lambda({a_0}_{(2)}...{a_n}_{(2)})
=\eta(a_0,...,a_n)\lambda(1)$ for all bounded linear functionals $\lambda$ on
$\cls.$  Similarly the case of the even chern characters can be treated.\\

\bcrlre
 Let us consider the case $\cla=\cls$, $\cla^\infty=\cls^\infty$, and
$\clh=L^2(h)$, with $V$ is the operator associated with the canonical regular
representationn of $\cls$ in $L^2(h)$, and let $R=F_{\phi_1}$  
 in $L^2(h)$. Given any positive operator $L$ on $L^2(h)$ such that
$L(t^n_{ij})=\lambda_n t^n_{ij}$ and $\sum_n M_n \lambda_n <\infty,$
 we can recover the Haar state by the following formula,\\
$$ h(a)=\frac{Tr(aRL)}{Tr(RL)}, a\in \cls.$$
\ecrlre
The proof of the corollary is immediate from the steps in the proof of the
Theorem \ref{main}. 

The above corllary generalizes a similar result obrained in \cite{Gos} for
$SU_q(2)$.  We have already remarked that for typical nonclassical
examples of compact matrix pseudogroups, we are forced to consider the
$\Theta$-summable case rather than the finitely summable case. Now, we shall
show that this assertion can be made a bit more definitive.
\bppsn
\label{negative}
Let $\cls$ be a compact matrix psedogroup with $\clh$ be $L^2(h)$ as before,
and suppose that the corresponding operator $R$ is not the identity, i.e.
$F_n$ is not equal to $I$. Assume that  there is an equivariant spectral
triple $(\cls^\infty, \clh, D)$ satisfying $[|D|,[|D|,a]] \in \clb(\clh)
\forall a \in \cls^\infty$. Then, there cannot be any finite positive number
$p$ such that $Lim_{t \raro 0+}(t^p  Tr(Re^{-tD^2}))=C , 0<C<\infty$, for some
suitable Banach limit $Lim$ on the space of bounded functions on $R_+$, as
considered in \cite{frolich} and elsewhere.  
 \eppsn
{\it Proof :-}\\
Suppose that the assertion of the proposition is false, and we are indeed
given an equivariant sepctral triple $(\cla^\infty, \clh, D)$ such that $[|D|,[|D|,a]] \in \clb(\clh)
\forall a \in \cls^\infty$ and $Lim_{t \raro 0+}t^pTr(Re^{-tD^2})=C,
0<C<\infty, p>0.$ 
Let $\eta(a)=Lim_{t
\raro 0+} \frac{Tr(aRe^{-tD^2})}{Tr(Re^{-tD^2})}$. 
We know from our earlier
results that $\eta(a)=h(a) \forall a \in \cls$. We  now claim that  
for $a,b \in \cls^\infty$, $\eta(ab)=\eta(\sigma(b)a)$, where
$\sigma(b)=\phi_1 \ast b$.  Before we prove this claim, we argue how it leads
to a contradiction and hence completes the proof. It is known from \cite{Wo1}
that $h(ab)=h((\phi_1 \ast b \ast \phi_1)a)$, where $(b \ast \phi):=(\phi \ot
id)(\Delta(b))$. Thus, we get that $(\sigma(b) \ast \phi_1)=\sigma(b) \forall
b \in \cls^\infty$, and as $\sigma$ is an automorphism, $b \ast \phi_1=b$. But
this is possible only if $F_n=I \forall n$, which is contradictory to the
assumption.

So, it is enough to prove that $\eta(ab)=\eta(\sigma(b)a)$. We have, 
\bean 
\lefteqn{R[e^{-tD^2},a]}\\ 
&=& -t \int_0^1 Re^{-tsD^2}[D^2,a]e^{-t(1-s)D^2}ds\\ 
 &=&-t \int_0^1 Re^{-tsD^2}(2|D|[|D|,a]+[|D|,[|D|,a]])e^{-t(1-s)D^2}ds. 
\eean   
By standard estimates one can now show that $t^p $ times the above expression
goes to $0$ in trace-norm as $t \raro 0+$, and from this the claim is
verified. 

\brmrk
\label{choiceofr}
If we look at the proof of the Theorem \ref{main} carefully, we can easily
notice that there is indeed some amount of flexibility in the choice of $R$.
In fact, the conclusion of the theorem will be valid if we replace the
canonical $R$ chosen by us by some operator of the form $R_1=RR^\prime$, where
$R^\prime$ is a positive operator having $e^{n,k}_j$'s as a complete set of
eigenvectors with $R^\prime(e^{n,k}_j)=\mu_k e^{n,k}_j,$ with $\mu_k$'s be
such that $R_1e^{-\beta D^2}$ is trace-class for all $\beta >0.$ In the
context of $SU_q(2)$, this flexibility of choice will play a crucial role.
However, it should be noted that the conclusion of Proposition \ref{negative}
does no longer hold if we change $R$.
  \ermrk

\brmrk
 The twisted Chern character can be paired with the
equivariant $K$-theory, i.e. the $K$-theory of the subalgebra $\cla_{inv}
\equiv \{ a \in \cla : \delta(a)=a \ot 1_{\cls} \}.$ In fact, from the special
from of $\sigma$, it is easily seen that $\cla_{inv} \subseteq \cla_\sigma,$
 hence we can restrict the pairing $<.,.>_{\sigma, \epsilon}$ on
$K_*(\cla_{inv}) \times H^*_\epsilon (\cla, \sigma)$ to get the desired map.
Furthermore, since the equivariant Dirac operator $D$ decomposes into a direct
sum of operators $D_\pi$, indexed by irreducible representations $\pi$ of
$\cls$, and any projection of $\cla_{inv}$ also naturally respects this
decomposition, one can consider the twisted Chern character corresponding     
 to the spectral data given by any fixed $P_\pi D P_\pi$, where $P_\pi$
denotes the projection onto the subspace corresponding to $\pi$. The
corresponding pairing assigns to each element of $K_*(\cla_{inv})$ a complex
number depending on $\pi$, and thus one gets a map from the set of irreducible
representations of $\cls$ to the dual of the $K$-theory, which 
may be formally thought of  some kind of ``character-valued index". However,
any attempt to give this a rigorous meaning requires first of all a
 generalization of equivariant entire cyclic cohomology as discussed, for
example, in \cite{KL} and related works of other authors, to the framework of
compact quantum groups. To the best of our knowledge, such a generalization
has not yet been achieved.
   \ermrk

We shall conclude with some discussion on the case of $SU_q(2)$. 
We recall from section 2 that in general the twisted entire cyclic cohomology
$H^*_\epsilon(\cla, \sigma)$ pairs with the $K$-theory of only a subalgebra
$\cla_\sigma$ of $\cla$, and not with that of $\cla$. However, it may
sometimes  turn out that the subalgebra $\cla_\sigma$ is large enough to
capture the $K$-theory of $\cla$ itself. We shall see that this is indeed the
case if we consider $SU_q(2).$ Let us recall the notation of \cite{CP}, where
the generators of $SU_q(2)$ were denoted by $\alpha, \beta$, and
$u=I_{1}(\beta^*\beta)(\beta-I)+I$ was choesn to be the generator of
$K_1(SU_q(2))$ which is $Z.$ Since the spectrum $\sigma(u)$ is a connected set
(which can be easily verified from the litarature on $SU_q(2)$), and $u$ is
invertible, it is clear that the map from $K_1(C^*(u))$ to $K_1(SU_q(2))$,
induced by the inclusion map, is an isomorphism of the $K_1$-groups (where
$C^*(u)$ denotes the unital $C^*$-algebra generated by $u$).     Now our aim is
to construct an appropriate twisted spectral triple so that the associated 
fixed point subalgebra $SU_q(2)_\sigma$ will contain $u$. To do this, we have
to refer to the remark \ref{choiceofr} made   earlier. We consider any of
the equivariant spectral triple constructed by the authors of \cite{CP}
and in the associated Hilbert space (which is the canonical regular
representation space of $SU_q(2)$) choose $R^\prime$ to be the operator with
corresponding eigenvalues  $\mu_k=q^{-2k},$ so that the new choice of $R$
actually coincides with that in \cite{Gos}. It can be easily verified that for
$\sigma$ corresponding to this choice of $R$, the fixed point subalgebra
$SU_q(2)_\sigma$ is the unital $\ast$-algebra generated by $\beta$, so in
particular contians $u$, allowing us to consider the pairing of the twisted
Chern character with $K_1(C^*(u))$, and in turn with $K_1(SU_q(2))$ using the
isomorphism noted before. The important question is whether we recover the
pairing obtained in \cite{CP} in our twisted framework, or if   our
pairing nontrvial. We conjecture that this question has an affirmative answer,
but to calculate the pairing we shall need to build some more tools, analogous
to the index theorem available for the untwisted or conventional set-up. We
however would like to postpone these issues for future works.

\end{document}